\documentclass[reprint, showpacs, amsmath,  amsfonts, prb]{revtex4-1}
\usepackage{graphicx}
\usepackage{amsmath}
  \usepackage{amssymb}
  \usepackage{bm}

\usepackage[cp1251]{inputenc}
\usepackage[T1,T2A]{fontenc}

\usepackage{color}

\begin{document}
\date{\today}
  \title{Application of the generalized Kirchhoff's law to calculation\\ 
of photoluminescence spectra of one-dimensional photonic crystals}

   \author{M. M. Voronov}
   \email{mikle.voronov@coherent.ioffe.ru}
    \affiliation{Ioffe Institute, St. Petersburg 194021, Russia}

\begin{abstract}
The approach based on the generalized Kirchhoff's law for calculating 
photoluminescence (PL) spectra of one-dimensional (1D) multi-layered
structures, in particular, 1D photonic crystals has been developed. It is
valid in the local thermodynamic equilibrium approximation and leads to 
simple and explicit expression for the photoluminescence intensity. In 
the framework of the present theory the analytical expression for the
spontaneous emission intensity enhancement factor (IEF) for a 1D photonic
crystal has been derived. It takes a particularly simple form in the case
of a sufficiently large number of the layers and is well suitable for 
analysis; in particular, it explains the difference in emission 
intensity at frequencies near different edges of photonic band-gaps 
(PBGs), where the intensity is relatively high, and specificity of 
suppression of the emission in a given frequency range. Also, the 
developed approach is discussed in connection with the standard method 
using the Fermi's golden rule and the concept of the local density of 
states (LDOS).
\end{abstract}

 \maketitle
 
\section{Introduction}

The theory of light emission from microstructures has been described in 
many books and reviews, see, e.g., [\onlinecite{Busch,Novotny}]. The 
principle difference in the calculation methods is related to the 
electromagnetic field quantization, which in most experiments on 
luminescence is not apparent. This situation corresponds to the so-called 
weak-coupling regime where the atom-cavity coupling constant, $g$, is much 
less than the cavity decay rate, $\kappa$, and the dipole decay rate, 
$\gamma$, which are due to two energy-loss mechanisms. The opposite case
where the dipole-cavity interaction dominates  over the dissipative 
processes and zero-point energy must be taken into account is the 
strong-coupling regime [\onlinecite{Agarwal}]. It corresponds to the 
condition $g\gtrsim\kappa$,$\gamma$ and is realized in high quality-factor 
microcavities and photonic crystals within the photonic band-gap region 
making spontaneous emission a reversible process and in the case of 
$g>>\kappa$,$\gamma$ (the quantum-coherent coupling regime) leading to 
the vacuum Rabi oscillations [\onlinecite{Milonni}]. The weak-coupling regime is 
characterized by an exponential decay of the light intensity, but with a 
different decay rate compared to that in free space (the Purcell effect 
[\onlinecite{Purcell}]), which is perfectly realized, for instance, for a 
1D photonic crystal with a moderate number of periods. 
This is just the situation considered in the present paper.

In this paper we present a theory of photoluminescence of one-dimensional 
layered structures on the basis of the generalized form of Kirchhoff's law. 
It determines the spectral dependence of the photoluminescence intensity 
when the lifetime of excited states ($\tau$) of emitting atoms, 
which we will term
\textquotedblleft emitting centers\textquotedblright\: (ECs) or simply 
\textquotedblleft emitters\textquotedblright, is long enough compared to 
the thermalization time ($\tau_{th}$) of electrons participating in optical transitions 
and short compared to the energy redistribution time for the rest of the system, 
so that the energy distribution stays the same while the spectrum is being measured.
In this case, which is typical of steady-state luminescence experiments, it is 
possible to use the concept of local thermodynamic equilibrium (LTE) and introduce
a local temperature to describe the quasi-equilibrium distribution
[\onlinecite{Sharkov}].
(It is supposed that 
the local temperature doesn't change during the measurement time.)
As the condition of quasi-thermal equilibrium is sufficient for the generalized 
Kirchhoff's law to hold true, it characterizes not only general 
thermodynamic properties of the light-emitting systems but also the 
specificity of the emitting material. A similar approach is used to 
quantitatively describe luminescence spectra of very different systems 
including astrophysical objects [\onlinecite{Wu:rfel,Ries,Ferraioli,Sharkov}]
and is sometimes mentioned in the literature as the generalized Planck's law 
[\onlinecite{Henneberger,Daub}]. Also, there exists a well-known relation
between the absorption and luminescence spectra called the van 
Roosbroeck–Shockley relation [\onlinecite{Roosbroeck}] (and also the 
Kennard-Stepanov relation [\onlinecite{Kennard,Stepanov}]).

As is known, in the semi-classical theory of radiation the atom is treated as
a quantum mechanical system while the electromagnetic field is described 
classically. However, to get the true value of the Einstein A coefficient for
spontaneous emission one has to take into account both the radiation reaction
field and vacuum field fluctuations [\onlinecite{Milonni}]. Since the type of 
emitters in our theory plays no important role, this theory can be considered 
equally as classical and as semi-classical and, consequently, the question 
about the origin of spontaneous emission does not arise here.
The emitting centers are efficient sources of spontaneous emission and, 
as is discussed in the paper, the Langevin-like approach
in which the sources terms are contained in Maxwell’s equations
[\onlinecite{Deych}] is equivalent to the generalized Kirchhoff's law under study.

The analytical expressions presented in this paper can be used in calculations
of the PL spectra for multi-layered structures (which in many experimental
situations can be considered as quasi-one-dimensional), in particular, for a
1D photonic crystal composed of the layers uniformly doped with the emitting 
centers. The numerical calculation procedure using these expressions is much
less cumbersome and more explicit compared to a commonly used method based on
the Fermi's golden rule approach and the concept of the local density of 
optical states derived from the Green's function of the system 
[\onlinecite{Novotny,Sprik,Vos}]. The last method, as is shown in the
paper, in the case of low quantum efficiency leads to the same answer as 
the approach developed here. Its application is 
illustrated on the example of calculation of the IEF for a 1D photonic 
crystal composed of two types of layers for the case when the light emission 
is generated from only one type of layers. It is possible in this case to make
a relatively simple analysis which helps reveal some features of the IEF such
as its asymmetry about the center of a PBG and the evolution of the emission 
peaks with a change in the number of periods of the structure. On the whole, 
this theory is a development of the so-called \textquotedblleft
indirect\textquotedblright method presented in [\onlinecite{Voronov}] for the case of
quasi-thermal equilibrium and differs from the latter and other theories of the
emission in periodic structures in giving an analytical expression for the 
emission intensity.

\section{The generalized Kirchhoff's law for 1D layered structure}

As a model system, we consider a 1D layered structure consisting of 
alternating plane layers, A and B, with real values of the 
refractive indices $n_a$ and $n_b$, respectively. The active (A) layers 
contain a large number of emitting centers (e.g., complexes of impurity atoms 
or ions with the nearest neighbour ions and also defects of the crystalline structure),
differing, in general, in their 
rates of radiative (and non-radiative) transitions; we assume for simplicity 
that the concentration of emitting centers in each A layer is the same 
and is not too large so that the effects of their interaction with one another
can be neglected (the approximation of independent emission sources).
In the present theory, the specific nature of such emitters is of no importance from the 
point of view of the general approach and is taken into account separately. 
It is partly due to the fact that the underlying Kirchhoff's law of thermal 
radiation is independent of the equilibrium-producing material and it is valid 
if the energy distribution of the excited states of the emitting centers
and, as a consequence, the shape of the resulting PL emission spectrum 
(in a homogeneous bulk material), is independent of frequency of the exciting laser beam; 
it is typical for thermalized photoluminescence, which is the case considered
in this study.

We start with the equation for the photoluminescence intensity for a 
one-layer structure [\onlinecite{Voronov,Deych}]:
\begin{equation}
I(\omega)\propto\omega N_{ph}(\omega)A'(\omega)\:.
\end{equation}
Here $N_{ph}(\omega)$ is the photon distribution function (which is 
approximately $\exp(-\hbar\omega/k_BT)$ at not too high temperatures,
$k_BT\ll \hbar \omega$) and $A'(\omega)$ is the absorption 
coefficient: $$A'(\omega)=1-|r(\omega)|^2-|t(\omega)|^2\:,$$ where 
$r(\omega)$ and $t(\omega)$ are the amplitude reflection and transmission 
coefficients of light at normal incidence; the prime on the function 
$A(\omega)$ denotes that absorption is very small, i.e. 
$A'(\omega)=\lim_{n''\rightarrow 0}A(\omega)$, where $n''$ is the imaginary 
part of the refractive index associated with the absorption of light by ECs. 
Hereafter we consider a layer-by-layer structure, so we define $A'_m(\omega)$ 
as the contribution to the absorption by the $m$-th A layer.
As follows from the excitation conditions for luminescence we 
should disregard the reabsorption and re-emission effects and thus to keep 
only the first-order term in $n''_{EC} (\omega)$ in the Maclaurin series for 
$A'_m(\omega)$, where $n''_{EC}(\omega)$ is proportional to concentration of
ECs, emitting at the frequency $\omega$. It follows then that $A'_m(\omega)$ 
can be represented by
\begin{equation}
A'_m(\omega) = \phi_m(\omega)n''_{EC}(\omega)\:,\:\: \phi_m(\omega)
= d A'_m(\omega)/d n''\:.
\end{equation}
Then the intensity of photoluminescence from the $m$-th A layer can be 
written in the form:
\begin{equation}I_m(\omega) \propto \omega
f_T(\omega)\phi_m(\omega)F_m(\omega_0)\:.
\end{equation}
This expression is obtained from Eqs. (1) and (2) by replacing 
$N_{ph}(\omega)n''_{EC}(\omega)$ with a new distribution 
function, $f_T(\omega)$, which depends on the local temperature $T$ of the 
system. 

Let us explain the meaning of the functions in Eq. (3). The function 
$f_T(\omega)$ is determined by the nature of ECs and the kind of the 
light-emitting material and thus is related to the bulk emission spectrum 
(in the absence of a quasi-standing wave resulting from the light reflection
from two interfaces). The function $\phi_m(\omega)$ is determined 
exclusively in terms of parameters of the structure (in particular, it is 
dependent on the total number of layers) and is responsible for modification
of the spectrum compared to that from a bulk material. The function 
$F_m(\omega_0)$ in Eq. (3) gives the relative intensity of the light 
absorbed in the $m$-th A layer at the excitation frequency $\omega_0$. 
The expression for the function $F_m(\omega)$ and its connection to $\phi_m(\omega)$ 
for a 1D photonic crystal will be given below. 

The photoluminescence intensity $I_N(\omega)$ from the structure containing 
$N$ active (A) layers is the sum of contributions:  
$I_N(\omega)=\sum_{m=1}^N I_m(\omega)\:.$ In the case of a small 
value of the imaginary part of refractive index of the constituent 
materials at the frequency $\omega$, one can set 
$A'_N(\omega) = \sum_m A'_m(\omega)$. Hence 
$\Phi_N(\omega) = \sum_m \phi_m(\omega)$ and, consequently, by neglecting 
the difference in values of $F_m(\omega_0)$ for different A layers, one gets
\begin{equation}
I_N(\omega) =\hbar \omega f_T(\omega)\Phi_N(\omega)\:,
\end{equation}
which as well as Eq. (3) we call the generalized Kirchhoff's law for a 
one-dimensional layered structure. The equality sign in Eq. (4) means only
that the distribution function $f_T(\omega)$ is appropriately normalized.
In essence, Eqs. (3) and (4) represent a modified form 
(suitable for a layered solid-state structure) of the Kirchhoff's law 
in the theory of radiation transfer, which expresses the equality
between the directional spectral emissivity and absorptivity
for non-polarized radiation [\onlinecite{Sharkov}]. 
It is valid in the stationary case at LTE condition  
even when the local temperature changes in space, $T=T(z)$ [\onlinecite{Sharkov}]. 
(In our study, for simplicity and without loss of generality, we assume that 
$f_T(\omega)$ is independent of $z$.) Thus, the generalized Kirchhoff's law 
is applicable to nearly transparent media and is not suitable, for instance, for 
microcavities with metallic walls (though the weak-coupling regime is 
fulfilled). 

It is worth mentioning that if the emitting centers are contained not 
only in the A layers but also in the B layers, the calculation of 
emission intensity should be made by taking into account the 
contributions from both types of layers on the basis of
the following expression:
\begin{equation}
A'(\omega)=n''_a\Phi^{(a)}_N(\omega)+n''_b\tilde{\Phi}^{(b)}_{\tilde{N}}
(\omega)\:,
\end{equation}
where $N$ and $\tilde{N}$ are the numbers of the A and B layers, 
respectively, and
$$\Phi^{(a)}_N (\omega)=\left(\frac{\partial A(\omega)}{\partial n''_a}\right)
_{n''_b=0},\:\:\:\tilde{\Phi}^{(b)}_{\tilde{N}} (\omega)=\left(\frac{\partial A(\omega)}
{\partial n''_b}\right)_{n''_a=0}\:.$$
The function $\Phi^{(a)}_N (\omega)$ should be calculated in the limit
$n''_a\rightarrow 0$ and it may be written in analogy with Eq. (2) as 
$\sum_m \phi^{(a)}_m(\omega)$, where the sum is over all of the A layers
in the structure. The function $\tilde{\Phi}^{(b)}_{\tilde{N}}(\omega)$ is 
defined in an analogous way. With the above taken into consideration, Eqs. 
(3) and (4) can easily be generalized to the case of two types of active 
layers, A and B. In the following sections, for simplicity, we confine 
ourselves to the simplest case when only the A layers are luminescent 
(so that $n''_b=0$) and for the sake of shortness, instead of 
$\Phi^{(a)}_N(\omega)$, we will use the notation $\Phi_N (\omega)$, as
earlier. As shown below, the function $\Phi_N (\omega)$ is related to
the spatial distribution of the energy density and, as a consequence,
to the emission intensity outside the structure, 
therefore we will call it the photoluminescence (PL) spectral function.

\section{A general calculation of photoluminescence intensity for 1D 
photonic crystal}

Now we will describe the calculation procedure of the photoluminescence
intensity from a 1D photonic crystal. We shall consider the
case of normal incidence of light on the structure from a medium with a 
refractive index $n_b$. It is useful to give the expressions for the 
amplitude reflection and transmission coefficients for the structure with
an arbitrary number $N$ of the A layers [\onlinecite{Ivchenko}]:
$$
r_N =r_1 \sin{NQd}/Z_N,\:\:t_N =t_1\sin{Qd}/Z_N\:,
$$
where $r_1$ and $t_1$ are the reflection and transmission coefficients for
a single A layer.
The other notations used are: \\ $\varphi_a = \omega n_a a/c,\:
\varphi_b = \omega n_b b/c, \:r = (n_a-n_b)/(n_a+n_b), 
\\t = 2n_b/(n_a+n_b),$ 
$\xi = 1-r^2 e^{2i \varphi_a}$ and 
\begin{equation}
Z_N (\omega) = \sin{NQd} - t_1 \sin{(N-1)Qd}\:,
\end{equation}
where $a$ and $b$ are the thicknesses of the A and B layers, respectively,
and $Q$ is the magnitude of the wave vector, which satisfies the equation
\begin{equation}
\cos Qd=\cos\varphi_a\cos\varphi_b - \frac{1}{2}\left(\frac{n_a}{n_b}+
\frac{n_b}{n_a}\right)\sin\varphi_a\sin\varphi_b\:.
\end{equation}

It is convenient to introduce $\eta_m = t_m/(1-r_m r_{N-m})$. 
By considering the energy flux density through the $m$-th A layer with the 
imaginary part of the refractive index $n''$, one can show that the 
function $F_m(\omega)$, which in the limit of $n''\rightarrow 0$ turns to
$F'_m(\omega)\equiv \phi_m(\omega)n''$, can be written in the form
\begin{equation}
F_m (\omega) = |\eta_{m-1}|^2 (1-|r_{N-m+1}|^2) - |\eta_m|^2 (1-|r_{N-m}|^2)\:.
\end{equation} This expression can be used for approximate calculation of
the intensity of the pump, $F_m(\omega_0)$, at an arbitrary value of 
$n''(\omega_0)$. (The prime on the function $F_m(\omega)$ implies that 
$n''\rightarrow 0$.)

Let us introduce the ratio 
$\gamma_m(\omega)\equiv I_m(\omega)/I_1^{(0)}(\omega)$, where 
$I_1^{(0)}(\omega)$ is the intensity from a single A layer of the thickness 
$a$, if the surrounding medium has the same refractive index as the A layer, 
$n_b=n_a$. Making use of Eq. (2) at $m$=1, we find that
in the absence of dielectric contrast the function $\Phi_1(\omega)\equiv\phi_1(\omega)$ is
$\Phi^{(0)}_1(\omega) = 2\omega a/c$. From Eq. (3) without taking into 
account $F_m(\omega_0)$ one gets 
\begin{equation}
\gamma_m(\omega)=c\phi_m(\omega)/(2a\omega)
\end{equation}
and the analogous quantity for the whole structure 
\begin{equation}
\Gamma_N(\omega)\equiv I_N(\omega)/(NI_1^{(0)}(\omega)) = c\Phi_N(\omega)/(2Na\omega)\:,
\end{equation}
which we call the spontaneous emission intensity enhancement factors. 
Evidently, the functions $\gamma_m(\omega)$ and $\Gamma_N(\omega)$ give 
the relative change in emission intensity for the $m$-th A layer and 
$N$-period structure, respectively. Hence, together with Eq. (4), one 
obtains the expression 
$I_N(\omega)\propto \omega^2f_T(\omega)aN\Gamma_N(\omega)$.
The function $\Gamma_N(\omega)$, as well as $\Phi_N(\omega)$, is 
expressed only in terms of the parameters of the photonic crystal and does
not depend on the emission characteristics of the sources (the emitting 
centers).  It is responsible for modification of the emission spectrum 
due to the dielectric environment of the emitting centers (because of 
numerous reflections of the light when it is propagating in the structure), 
while the function $f_T(\omega)$ is determined by the population of states
of the centers and is proportional to their concentration.
The calculation of the function $f_T(\omega)$ is a separate problem; 
however, as follows from the above analysis, when the distribution of 
nonequilibrium carriers between the states can be described by means 
of quasi-Fermi levels for electrons and holes, 
the function $f_T(\omega)$ contains the absorption coefficient $\alpha(\omega)$ 
as one of the multipliers.
In a bulk sample (a dielectrically homogeneous medium), as $\Gamma_N(\omega)=1$, 
the PL intensity is proportional to
$\omega^2f_T(\omega)$.
Notice that to get the maximum value of the intensity of 
photoluminescence, $I_N(\omega)$, the frequency corresponding to the 
highest peak of the function $\Gamma_N(\omega)$ should coincide with the 
frequency of the emission spectrum maximum of the bulk material.

Using the generalized Kirchhoff's law, one can make a more exact calculation
of the photoluminescence intensity, by taking into account the difference in
the light absorption in different regions of the $m$-th active layer. We now
note that the function $\phi_m(\omega)$ corresponds to the time-average 
power of the (monochromatic) electromagnetic field absorbed per unit volume 
in the $m$-th active layer, 
$P=(1/8\pi)\omega\varepsilon''|{\bf E}(\omega,{\bf r})|^2$,
(see [\onlinecite{Landafshiz}]) integrated over the thickness of the layer; 
in our case $\varepsilon'' = 2n_a n''$. For the structure under consideration,
which is translationally invariant in x and y, one can represent
$E(\omega, {\bf r})=e^{i{\bf q}{\bf \rho}}{\bf\cal{E}}_q(\omega,z)$,
where $\bf{\rho}$ and $\bf{q}$ are the in-plane vectors. In the case when a plane 
electromagnetic wave of the frequency $\omega$ is normally incident on the 
structure ($q=0$), one can write
\begin{equation}\label{phi/field}
\phi_m(\omega)=C\omega\int_{-a/2}^{a/2}|{\cal{E}}^{(m)}(\omega,z)|^2 dz\:,
\end{equation}
where the function ${\cal{E}}^{(m)}(\omega, z)$ describes the 
distribution of the electric field along the $m$-th active layer and $C$ is
a coefficient. Applying the generalized Kirchhoff's law to an 
infinitesimally thin layer and taking into account the last relation, after
integrating over the thickness of the $m$-th A layer, we arrive at the 
following relationship:
\begin{equation}
I_m(\omega) \propto \omega^2 f_T(\omega)\int
dz|{\cal{E}}^{(m)}(\omega_0,z)|^2 |{\cal{E}}^{(m)}(\omega,z)|^2\:.
\end{equation}
Note that if the local temperature depends on $z$ alone, so that 
the temperature distribution $T=T(z)$ is smooth and constant in time,
the function $f_T(\omega,z)$ should be inserted into the integrand.
As $f_T(\omega,z)$ is proportional to the concentration of ECs,
it will also be included under the integral if the concentration changes 
with $z$.

The electric field ${\cal{E}}^{(m)}(\omega,z)$ can be written as
\begin{equation}
{\cal{E}}^{(m)}(\omega,z)={\cal F}^{(m)}_1(\omega) e^{ik_a z}
+ {\cal F}^{(m)}_2(\omega) e^{-ik_a z}\:,
\end{equation}
where the $z$-coordinate is measured from the center of the $m$-th A layer,
$k_a=\omega n_a/c$ 
and the functions ${\cal F}^{(m)}_1$ and ${\cal F}^{(m)}_2$ are
${\cal F}^{(m)}_1=e^{i(\varphi_a +
\varphi_b)/2}(\eta_{m-1}+\eta_{m}r_{N-m}re^{i\varphi_a})/\xi\:,$\\
${\cal F}^{(m)}_2=e^{i(\varphi_a +
\varphi_b)/2}(\eta_{m-1}re^{i\varphi_a}+\eta_{m}r_{N-m})/\xi \:.$
The function $|{\cal{E}}^{(m)}(\omega_0,z)|^2$ is responsible for excitation
of photoluminescence by incident light with a frequency $\omega_0$. 

The simplest way to get the coefficient of proportionality $C$ is to compare 
the function $\Phi_1(\omega)$  with  
Eq. (11) at $m=1$; hence $C=(2/c)(1-r^2)$, where $c$ is the 
speed of light in vacuum. The calculation of $\phi_m(\omega)$ with using Eqs.
(11) and (13) leads to 
\begin{eqnarray}
\phi_m(\omega) &=& \frac{2(1-r^2)\omega a}{c|\xi|^2}[g_1 (|\eta_{m-1}|^2+
|\eta_m r_{N-m}|^2) + \nonumber\\ 
& & 2g_2Re(\eta_m\eta^*_{m-1}r_{N-m})]\:,
\end{eqnarray}
$$g_1=1+r^2+r\frac{\sin2\varphi_a}{\varphi_a}\:,\:\:\:g_2=(1+r^2)
\frac{\sin\varphi_a}{\varphi_a}+2r\cos\varphi_a\:.$$

One can show that the calculation procedure described above is equivalent to
the calculation of the light power generated by independent sources. (It is
in accordance with the theory presented above, where the reabsorption and 
re-emission processes were ignored.) For this purpose one should express the
intensity of radiation transmitted through the photonic crystal in terms of 
the amplitude of a plane electromagnetic wave coming from the plane 
$z=const$ (where $z$-axis is defined inside the $m$-th active layer with 
the origin in the layer center) and calculate the reflection and transmission
coefficients taking into account all reflections and all re-reflections of
electromagnetic waves from various interfaces of the layers of the structure.
The resulting expression will have the same form as Eq. (12) in which
${\cal{E}}^{(m)}(\omega_0,z)=const$ and with ${\cal{E}}^{(m)}(\omega,z)$ 
proportional to Eq. (13). In a more rigorous consideration,
one has to resort to a Langevin-like approach,
where the wave equation (for the electric field
$\bf{E}(\omega, {\bf r})$ given above) must be solved 
taking into account the polarization sources, which are due to the
incoherent nature of spontaneous emission and can be
described by random functions of the coordinates and time.
This method is demonstrated in [\onlinecite{Deych}] 
on the example of a multiple-quantum-well structure, in which case 
the random term is ascribed to the exciton polarization.
Also, this method was used for calculations of steady-state photoluminescence spectra
of Fibonacci photonic quasicrystal containing organic dye molecules
[\onlinecite{Passias}].

In practice to calculate the PL intensity $I_N(\omega)$ of the structure with
$N$ active layers, one uses the photoluminescence spectrum 
$I^{(0)}_1(\omega)$ of a single active layer (of the thickness $a_0$). As 
follows from Eq. (12), the photoluminescence intensity under excitation at 
frequency $\omega_0$ can be estimated as
\begin{eqnarray}
I_N(\omega)=&&
I^{(0)}_1(\omega)\sum^N_{m=1}\int_{-a/2}^{a/2}J^{(m)}(\omega_0,\omega,z)dz\nonumber\\
&&\times\left(\int_{-a_0/2}^{a_0/2}J^{(1)}(\omega_0,\omega,z)dz\right)^{-1}\:,
\end{eqnarray}
where 
$J^{(m)}(\omega_0,\omega,z)=|{\cal{E}}^{(m)}(\omega_0,z){\cal{E}}^{(m)}(\omega,z)|^2$.
The equality sign in Eq. (15) implies that the power of excitation radiation
in the case of $N$-period structure is exactly equal to that for the
one-layer structure, otherwise it should be replaced by the sign of 
proportionality. If a 1D photonic crystal is terminated from one side with a plane interface 
between the material B and a medium with the refractive index $n_0$, so 
that the distance between two interfaces (of the three materials) is equal 
$b'$, Eq. (15) should be multiplied by the factor
$|\tau/(1-\varrho r_N e^{i\varphi_0})|^2$, where 
$\varrho = (n_b - n_0)/(n_b + n_0)$, $\tau = 2n_b/(n_b + n_0)$, 
$\varphi_0 = (2b'-b)\omega n_b /c$, and the electric 
field ${\cal{E}}^{(m)}(\omega_0,z)$ should be calculated taking this medium into account.

\section{The photoluminescence spectral function and intensity
enhancement factor of 1D photonic crystal}

As is seen from Eq. (4) the main features of the luminescence spectra of 
1D photonic crystal can be established from an analysis of the PL spectral
function $\Phi_N(\omega)$ or, as discussed above, the emission intensity
enhancement factor  $\Gamma_N(\omega)$, see Eq. (10). Taking the sum over 
$m$ from 1 to $N$ on both sides of Eq. (2)  and then calculating the 
derivative of $A'_N(\omega)$ one gets the following expression for the 
function $\Phi_N(\omega)$:
\begin{equation}
\Phi_N (\omega) = \frac{C_1 N + C_2 \sin
{NQd}}{(1-r^2)^2\sin^2{Qd}+(2r\sin\varphi_a)^2\sin^2NQd}\:,
\end{equation}
where the functions $C_1$ and $C_2$ are given by
\begin{eqnarray}
&&C_1 = [(1/n_b - n_b/n^{2}_a) \sin \varphi_a \sin \varphi_b \nonumber\\&&
+ (2\sin{ \varphi_a} \cos{ \varphi_b} + (n_a/n_b + n_b/n_a) \sin \varphi_b
\cos \varphi_a)\omega a/c ] \nonumber\\&& (1-r^2)[r^2 \sin
(\varphi_a-\varphi_b) + \sin (\varphi_a+\varphi_b)]\:, \nonumber
\end{eqnarray}
\begin{eqnarray}
&& C_2 = B_0 \sin{NQd} + B_1 \sin{(N-1)Qd} \nonumber \\ && +\ B_2
\sin{(N-2)Qd} - C_1 (\sin{NQd} + \cos{NQd} \cot{Qd})\:, \nonumber
\end{eqnarray}
\[
B_0 = 4[rt\sin 2\varphi_a /(n_a + n_b) + r^2 (1-r^2)(\omega a/c)]\:,
\]
\begin{eqnarray}
B_1 = -2(1-r^2)[3r^2 \cos (\varphi_b-\varphi_a) - 
\cos (\varphi_a + \varphi_b)](\omega a/c)\nonumber \\
-\ 8rt\sin \varphi_a \cos \varphi_b /(n_a + n_b)\:, \mbox{} \hspace{4 cm}
\mbox{} \nonumber
\end{eqnarray}
\[
B_2 = -2(1-r^2)^2 (\omega a/c)\:.
\]
It follows from Eq. (16)
that in the absence of dielectric contrast of the constituent materials 
(A and B) 
$\Phi_N^{(0)}(\omega)=2N\omega a/c=N\Phi_1^{(0)}(\omega)$.
In the long wavelength limit, at $\omega \rightarrow 0$, Eq. (16) converts
to $2Nan_a\omega/(cn_b)$.

In the most interesting case, when $N \gg 1$, Eq. (16) is greatly 
simplified and takes the form 
\begin{equation}
\Phi_N (\omega) \approx \frac{C_1(N-\sin NQd\cos NQd\cot{Qd})}{(1-r^2)^2\sin^2{Qd}
+(2r\sin\varphi_a)^2\sin^2NQd}\:.
\end{equation}
This expression is a good approximation in the frequency region where 
the values of $C_1(\omega)$ are not too small (see the case $C_1=0$ below).
A further simplification of the expression for $\Phi_N (\omega)$ can be 
made by setting in Eq. (16) $C_2=0$, which is valid for frequencies not 
too close to a PBG edge, where the condition $|\cot{Qd}\sin 2NQd|\ll 2N$ is 
satisfied; consequently, the numerator of Eq. (17) is equal to $C_1N$.
At the Brillouin zone center ($Q=0$) and at the Brillouin zone edge 
($Q=\pi/d$) the function $\Phi_N(\omega)$ becomes $\Phi_N^{(BZ)}=$
$$
\frac{2C_1N^3+3(B_0\pm B_1+B_2-C_1)N^2+(C_1\mp
3B_1-6B_2)N}{3[(2Nr\sin\varphi_a)^2+(1-r^2)^2]}
$$
where the upper sign refers to $Q=0$ and the lower sign refers to
$Q=\pi/d$. The expression for $\Phi_N^{(BZ)}$ shows that for a 
sufficiently large number $N$ of the layers ($N \gg 1$), approximately, 
$\Phi_N^{(BZ)} \propto N$.
\begin{figure}[b]
 \begin{center}
  \includegraphics[width=0.55\textwidth]{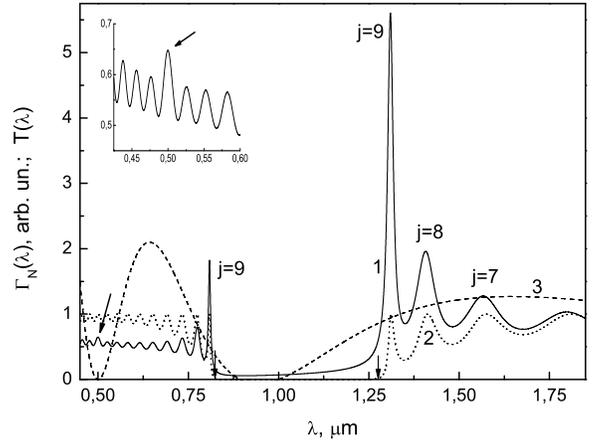}
 \end{center}
\caption{The intensity enhancement factor, $\Gamma_N(\lambda)$, (curve 1) 
and transmission coefficient, $T(\lambda)$, (curve 2) that has a correlation
with the IEF in the position of the spectral peaks. The graph of the
function $\Gamma_N(\lambda)$ is plotted by using Eqs. (10) and (16) and 
coincides with that plotted by using Eqs. (10) and (17) except for a small
region around $\lambda_c=0.5\:\mu m$, where Eq. (17) is not valid. The 
numbers $j=9,8,...$ in the figure indicate the corresponding spectral peaks. 
The curve 3 shows the function $C_1(\omega)$, which is equal to zero at
$\lambda_c=0.5\:\mu m$ (at which the second PBG collapses). The vertical 
arrows indicate the edges of the first PBG. The inclined arrow shows a
distinguished peak at the wavelength of $\lambda_c=0.5\:\mu m$, see also 
the inset of the figure. Calculated for a DBR with $N=10$ periods, the 
thicknesses $a={\bar\lambda}/(4n_a)=125$ nm, $b={\bar\lambda}/(4n_b)=250$ 
nm and refractive indices $n_a=2$, $n_b=1$ of the layers A and B, 
respectively; the tuning wavelength $\bar\lambda=1000$ nm.}
\end{figure}

As an example, Fig. 1 shows the IEF (curve 1) as a function of the
wavelength $\lambda$ for a distributed Bragg reflector (DBR) with $N=10$ 
active (A) layers, the transmission spectrum (curve 2) calculated in the 
absence of absorption and also the function $C_1(\lambda)$ (curve 3). As 
seen from the figure, curve 1 is essentially asymmetric and has a set of
peaks in two band regions, with the highest peaks being near the PBG 
edges. In the region of large wavelengths there are several spectral
peaks, which with increasing the wavelength become lower and wider; an 
analogous situation occurs in the region on the opposite side of the PBG.
Such a picture is typical of 1D photonic crystal; it is also observed when
the emitting centers are uniformly distributed in the B layers or in both 
types of layers. (Calculations are made using Eq. (5)). 

As follows from Eq. (16), the PL spectral function 
$\Phi_N(\omega)$ can be written as $P_N(\omega)/|\xi Z_N(\omega)|^2$, where 
$P_N(\omega)=C_1 N + C_2\sin {NQd}$, therefore the peaks of the function 
$\Phi_N(\omega)$ and, consequently, $\Gamma_N(\omega)$, are located around 
the real values, $\omega'_j$, of the complex eigenfrequencies 
$\omega_j=\omega'_j+i\omega''_j$ which can be found from the equation
$Z_N(\omega_j) = 0$, see Eq. (6). When absorption can be neglected the 
transmission coefficient in the band region is
$\:T(\omega)=|t_N(\omega)|^2=(1-r^2)^2\sin^2{Qd}/|\xi Z_N(\omega)|^2\:.$
As the denominator of the function $T(\omega)$ is the same as for 
$\Gamma_N(\omega)$, while the function $\sin^2Qd$ is rather smooth, the 
transmission peaks are located around $\omega'_j$ as well and are slightly 
shifted relative to the peaks of the function $\Gamma_N(\omega)$, see 
Fig. 1. Thus, the positions of peaks of the IEF, $\Gamma_N(\omega)$, are
close to those at which $T(\omega)=1$ and, consequently, the wave numbers 
corresponding to the local maxima of
$\Gamma_N(\omega)$ are $Q_j d=\pi j/N + \delta_j$, $|\delta_j| \ll 1$, 
where $j$ are integers and the frequencies $\omega_j$ satisfy Eq. (7) with 
$Q=Q_j$. Disregarding the value of $N\delta_j$, which for the peaks nearest
to the PBG is $|N\delta_j| \ll 1$, one gets from Eq. (17) 
\begin{equation}
\Phi_N(\omega_j)\approx\frac{C_1(\omega_j)N}{(1-r^2)^2\sin^2(\pi j/N)}\:.
\end {equation}
Upon moving away from a PBG edge and passing successive values of $\omega_j$,
the denominator (the squared sine) increases in a monotonous way and after 
that monotonically decreases when approaching to another PBG. A more exact 
consideration shows that when moving away from the PBG edge the values of 
$\delta_j$ can increase and should be taken into account, but qualitatively
the situation remains quite similar; since on the scale of the distance 
between the neighbouring peaks the function $C_1(\omega)$ changes relatively 
slowly compared to $\sin^2Q(\omega)d$, the peaks of the function 
$\Phi_N(\omega)$ and, consequently, $\Gamma_N(\omega)$ are getting lower.
The highest peaks correspond to $j=1$ (at the PBG edge $Q=0$) and $j=N-1$ 
(at the edge $Q=\pi/d$).  This explains why enhancement of the PL intensity 
occurs predominantly in the band region in the vicinity of the band-gap edge,
see, e.g., [\onlinecite{Pevtsov}]. It is obvious from Eq. (18) that if 
$j\ll N$ or $j\lesssim N$ (in the limit $N \gg 1$) 
$\Phi_N(\omega_j)\propto N^3$, so that with increasing the number of periods,
$N$, the peaks of the function $\Phi_N(\omega)$ increase and approach to the 
PBG edges, and their number increases as well (in accordance with Eq. (6) and
the simplified formula $Q_jd \approx \pi j/N$).

Another feature of the functions $\Phi_N(\omega)$ and $\Gamma_N(\omega)$ is 
their asymmetry relative to the PBG center. It can be easily explained with 
the help of Eq. (18) and expression for the function $C_1(\omega)$. Let us 
denote by $\omega_{j-}$ and $\omega_{j+}$ the frequencies corresponding to 
the peaks of the function $\Gamma_N(\omega)$ on different sides of a PBG. 
Evidently, $C_1(\omega_{j-})$ and $C_1(\omega_{j+})$ can take substantially 
different values that leads to an asymmetry in the height of the peaks with 
the same number $j$. In the example considered above (see Fig. 1) the PBG 
edges satisfy the condition $Qd=\pi$, consequently the peaks of 
$\Gamma_N(\omega)$ nearest to the PBG correspond to the integers $j=9,8,...$.
It follows from Eq. (18) that  
$\Phi_N(\omega_{j+})/\Phi_N(\omega_{j-}) \approx C_1(\omega_{j+})/C_1(\omega_{j-})$,
therefore the ratio 
$\zeta(j) \equiv \Gamma_N(\omega_{j+})/\Gamma_N(\omega_{j-})\approx \omega_{j-} 
C_1(\omega_{j+})/(\omega_{j+} C_1(\omega_{j-}))\equiv{\tilde\zeta}(j)$.
The calculation gives the following values: 
$\zeta(j=9) = 3.075$, ${\tilde{\zeta}}(j=9) = 3.045$ and $\zeta(j=8) = 2.439$, 
${\tilde{\zeta}}(j=8) = 2.408$, which verifies the applicability of Eq. (18).
The analysis of Eq. (17) allows one to determine approximate values of the 
local minima of the PL spectral function. They correspond to the wave numbers 
$Q_kd=\pi(2k+1)/(2N)$, where $k=1,...N-2$, hence $\Phi_N(\omega_k)\approx$
$$
\frac{C_1(\omega_k)N}{(1-r^2)^2\sin^2(\pi(2k+1)/(2N))+(2r\sin(\omega_kn_aa/c))^2}\:.
$$
This expression also explains the asymmetry of the PL spectral function and,
as a consequence, of the IEF.

The value of the function $C_1(\omega)$ changes considerably on the scale of
the distance between the edges of PBGs, therefore it essentially determines 
the value of the IEF at a given frequency, in particular for frequencies 
close to a PBG edge, where the values of $\Phi_N(\omega_j)$ and 
$\Gamma_N(\omega_j)$ can be large, see Eq. (18). Therefore, by choosing the 
appropriate parameters and, thus, \textquotedblleft governing\textquotedblright\:
the function $C_1(\omega)$ one can achive either an anomalously large enhancement
of the emission intensity or its moderate suppression. However, as follows from 
the analysis of the function $C_1(\omega)$, for frequencies in a photonic band
the inequality $C_1(\omega)>0$ is always satisfied. This means that in the case 
of a 1D photonic crystal (without a defect) the light emission in the regime 
$I_{N}\propto N$ at a PBG edge (because $\Phi_N^{(BZ)}\propto N$) and 
$I_{N}\propto N^3$ at the spectral peak frequency $\omega_j$ (because 
$\Phi_{N}\propto N^3$) cannot be suppressed completely, but can only be 
decreased due to a relatively small value of $C_1(\omega_j)$. In accordance with
Eq. (10), in these cases $\Gamma_N=const$ and $\Gamma_N\propto N^2$, where the 
latter corresponds to the superradiant regime, which is due to the periodicity 
of the structure. However, since in any physical system of the considered type 
the absorption of the emitted radiation takes place, as well as loss of coherence, 
for sufficiently large values of $N$ this quadratic dependence ceases to be valid;
moreover, in the case of very large $N$ and an extremely small absorption coefficient 
the high quality modes (in the vicinity of the PBG edges)
come into force, potentially leading to the strong-coupling regime, in
which case the present theory is not applicable.

There is a special case when at the frequency $\omega_c$  the photonic band-gap
vanishes and, as the analysis shows, $C_1(\omega_c)=0$. It is worthwhile to 
notice that at this frequency the function $\Phi_N(\omega)$ and, consequently, 
$\Gamma_N(\omega)$ has a local maximum. In the case considered in Fig.~1 the 
corresponding peak appears at the wavelength $\lambda_c=0.5\:\mu m$. 
This peak is higher than
the neighbouring ones, because at the wavelength $\lambda_c$ all of the 
functions $\phi_m(\omega)$, where $m=1,2,...N$, have local maxima (see inset of 
Fig. 2). A quantitative description of this effect can be made by using Eqs. 
(14) and (16). The qualitative explanation is that 
as the PBG is getting narrower two spectral peaks adjoining the PBG edges 
essentially evolve and at the frequency $\omega_c$ join to form a single peak.
Moreover, since in the case of a 1D photonic crystal the topology of the band 
structure for propagation of electromagnetic waves is determined by the number
of photonic band-gaps, one can expect that in energy spectrum at the point where 
the topology changes (at the frequency at which a PBG collapses) some 
peculiarity will arise. This, indeed, occurs and manifests itself as a local 
maximum of the PL spectral function $\Phi_N(\omega)$, which determines the power
of the electromagnetic radiation absorbed per unit volume. 
In general, this phenomenon is associated with the anomalous absorption of 
electromagnetic radiation in 1D photonic crystals, which occurs in the vicinity
of the frequency at which either the band-gap or allowed region collapses 
[\onlinecite{Vinogradov,Voronov2}].

\begin{figure}[t]
 \begin{center}
  \includegraphics[width=0.55\textwidth]{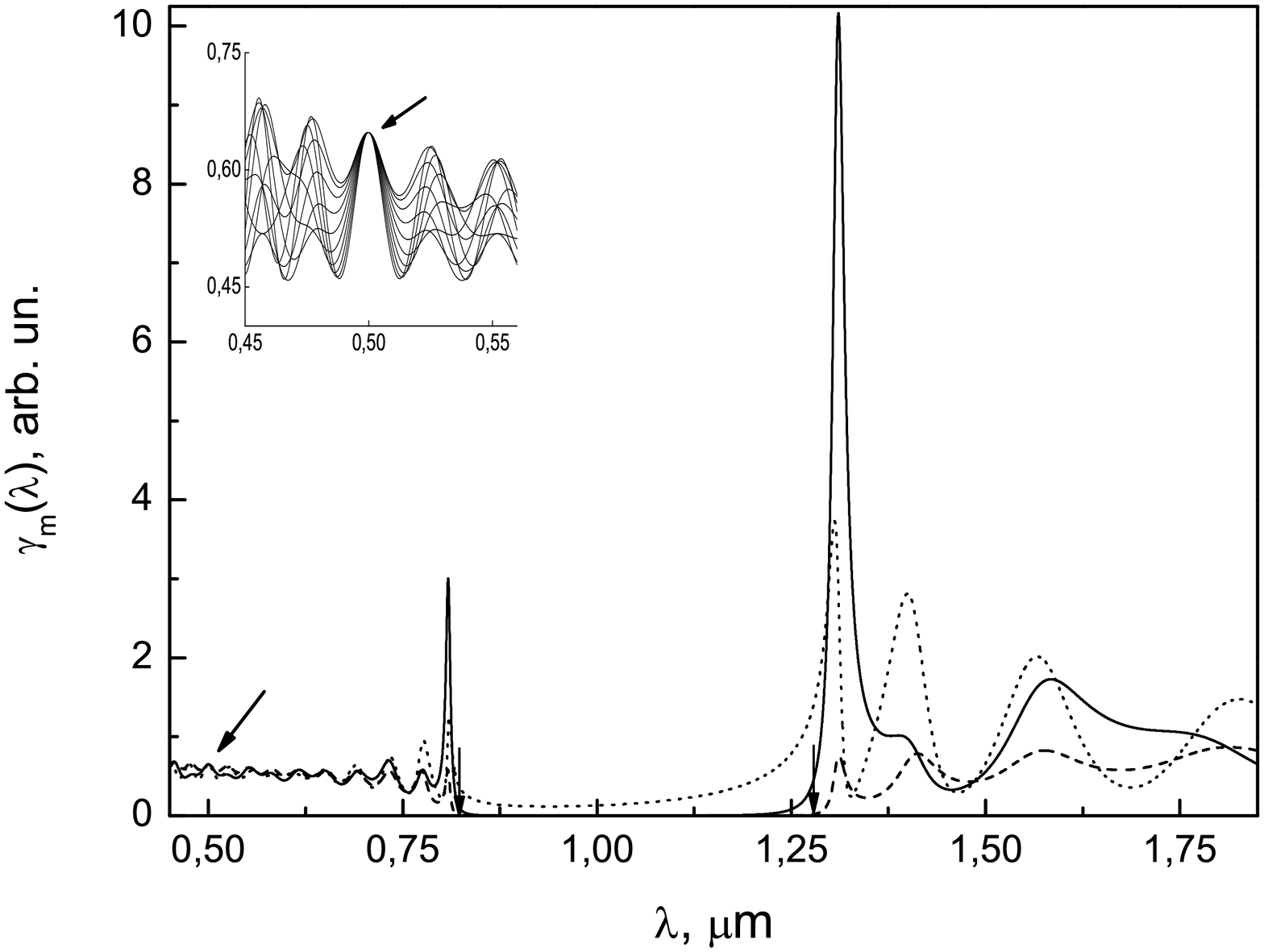}
 \end{center}
\caption{The partial IEFs, $\gamma_m(\lambda)$,
for the layers $m=2$ (dotted line), $m=6$ (solid 
line) and $m=10$ (dashed line). The vertical arrows indicate the PBG edges. The
inset shows a set of functions $\gamma_m(\omega)$, where $m=1,2,...10$, which 
at the wavelength $\lambda_c=0.5\:\mu m$ have the same local maximum value (the 
peak is indicated by arrow). Calculated for the same parameters as in Fig. 1.}
\end{figure}

We now briefly consider the behavior of the functions $\gamma_m(\omega)$, which
are the (partial) intensity enhancement factors for different emitting layers 
of the photonic crystal. Figure 2 shows the partial IEFs calculated by 
using  Eqs. (9) and (14) for the layers $m=2,6$ and 10 (see caption to Fig. 2). 
As seen from the figure, the central and first nearest layers give the largest
contributions to the total intensity (at different wavelengths), while the 
remote layers emit relatively weakly. This behavior of the functions 
$\gamma_m(\omega)$ can be explained from the analysis of Eq. (14). Qualitatively,
the electromagnetic radiation from a remote layer experiences strong reflection 
from the rest of the structure, while the transmittance of the radiation coming
from the first nearest layer at some wavelengths can be sufficiently high. The 
highest peaks of the functions $\gamma_m(\omega)$ near the PBG edges for the 
central layers ($m=$ 5 and 6) are due to some kind of microcavity effect, see Fig. 
2. In general, as shown in the previous section, the frequency dependence of the
IEF is associated with the distribution of electric field modes in the photonic 
crystal structure. 

\section{Discussion (connection to the Green's function method)}
Now we briefly discuss how the theory presented above is connected with the 
most commonly used theory of light emission from microstructures. As is known, 
there are two different ways to describe the emission process and calculate 
intensity of the emitted light, namely, classical and quantum approaches; their 
equivalence and difference have been discussed in detail in many papers and books,
e.g., see [\onlinecite{Novotny,Milonni,Loudon}]. In the classical approach the 
electromagnetic power is due to emitting dipoles and is given by
\begin{equation}
\frac{dW}{dt} \propto \omega^3|{\bf d}|^2\left<{\bf
e}_d\cdot Im[{\bf G}({\bf r},{\bf r},\omega)]\cdot{\bf
e}_d\right>\:,
\end{equation}
where ${\bf d}$ is the dipole moment oriented along the unit vector ${\bf e}_d$
and located at the  point $\bf r$, ${\bf G}({\bf r},{\bf r'},\omega)$ is a dyadic
Green's function (or Green's tensor) [\onlinecite{Tai}] of the considered system, and the angle 
brackets denote an average over all possible dipole orientations. In the dipole
approximation the emitting 
centers, just as atoms, are modelled as oscillating point dipoles 
having electric dipole moments.
In the case of uniform and isotropic distribution of dipole moments in an active (A) 
layer the average in Eq. (19) can be expressed through the trace of the imaginary 
part of the dyadic Green's function [\onlinecite{Novotny}].
In this connection we
study non-polarized radiation and consider the source (electric current density) 
and response (electric or magnetic field) as scalar functions; hence, the 
Green's function with which we will deal is a scalar function
corresponding to the symmetry of the system. This is in accordance with
the following qualitative consideration. Let us consider a
layer of (noninteracting) dipoles which is perpendicular
to the $z$-axis and whose thickness is much smaller than the light
wavelength. An ensemble of incoherent emitters with 3D isotropic random
orientation of the electric dipole moments gives
the same intensity as three incoherently radiating
dipoles with moments oriented along the $x$,$y$ and $z$ axes
[\onlinecite{Lukosz,Novotny2}]. Since we are interested in
calculating the far-field emission intensity (in the direction
perpendicular to the layered structure), one can align all
the dipoles in these three directions and then replace two-thirds of them with
a system of two sets of incoherently radiating infinite current sheets
with mutually perpendicular currents directed along
the $x$- and $y$-axes. Such replacement is possible because an
infinite plane sheet of dipoles directed in one direction and
oscillating in phase
radiates in the same way [\onlinecite{Sargent}] as an infinite plane
current sheet (which produces plane waves
propagating from both sides in the perpendicular direction),
while a set of $z$-directed dipoles gives no contribution to the
total intensity.
In the case of a current sheet the boundary conditions at the
interfaces of the layer are easily satisfied and as a result the
intensity of outgoing unpolarized radiation
will depend only on the $z$-coordinate of the current sheet in the
layer. Thus, by considering that dipoles are polarized in one plane, 
in the case of a one-layer structure one 
should take 1D scalar Green's function, calculate its imaginary part at the 
position of an emitting center for coincident emission and observation points,
$z=z'$, and then integrate over the layer thickness.

In the quantum approach in the case of the weak-coupling regime and low quantum 
efficiency [\onlinecite{Vos}] the rate of direct radiative transitions is given
by the Fermi's golden rule [\onlinecite{Loudon}]. As is shown in 
[\onlinecite{Sprik}], in this case the emission rate is proportional to the 
local density of states, $\rho({\bf r},{\bf e}_{\mu}, \omega)$, at the position of 
the emitting center: 
$\kappa_r(\omega)\propto \omega \mu^2\rho({\bf r},{\bf e}_{\mu},\omega)$,
where $\mu$ is the magnitude of the transition dipole moment and ${\bf e}_{\mu}$
defines its orientation. The LDOS in turn is proportional to the imaginary part
of the dyadic Green's function and calculated for the direction given by the 
orientation ${\bf e}_{\mu}$: 
$ \rho({\bf r},{\bf e}_{\mu},\omega) \propto \omega ({\bf e}_{\mu}\cdot Im[{\bf G}({\bf r},
{\bf r},\omega)]\cdot{\bf e}_{\mu})$. 
In the weak-coupling approximation the emission intensity can be 
calculated by the rate equation for the excited state population (the 
number of emitting centers in the excited state) and is given by 
[\onlinecite{Koenderink}] 
$I(\omega)\propto P\kappa_r(\omega)/(\kappa_r(\omega)+\kappa_{nr})$, where $P$ 
is the rate of excitation, and $\kappa_r(\omega)$ and $\kappa_{nr}(\omega)$ 
are the radiative and nonradiative decay rates for the transition from 
an excited state to the ground state. In the case of low quantum 
efficiency $\kappa_r\ll\kappa_{nr}$, so that 
$I(\omega)\propto P\kappa_r(\omega)/\kappa_{nr}$. 
Unlike the nonradiative decay rate $\kappa_{nr}$, 
which is mainly defined by the chemical composition of the A-layers and 
therefore can be taken constant, the quantities $P$ and $\kappa_r$ 
depend on the emitter position ($P=P({\bf r})$, $\kappa_r=\kappa_r({\bf r},\omega)$).
In the simplest case $P$ 
can be considered constant (otherwise it should be taken into account 
by an additional factor that is the squared magnitude of the electric 
field at frequency $\omega_0$ in the integrand of Eq. (12))
and, consequently, the spontaneous emission rate enhancement factor (the Purcell factor), 
$\gamma(\omega)\equiv \kappa_r(\omega)/\kappa_r^{(0)}(\omega)$, is 
approximately equal to $I({\bf r},\omega)/I^{(0)}({\bf r},\omega)$, where $I^{(0)}({\bf r},\omega)$
and $I({\bf r},\omega)$ are the emission intensities from a two-level 
quantum emitter in a homogeneous medium 
and from the emitter in the structure under study, 
respectively, and $\kappa_r^{(0)}(\omega)$ 
and $\kappa_r(\omega)$ are the corresponding radiative decay rates. 
As is known, if the atoms in a waveguide are initially in the excited state
they will spontaneously emit into a propagating mode (the light can be detected in the far field) 
and a guided mode (light does not leave the waveguide). The corresponding 
channels of the radiative decay contribute to the spontaneous emission rate 
and thus the both propagating and guided modes contribute into LDOS. 
In accordance with the above considerations about the replacement of the 
dyadic Green’s function by the 1D scalar Green's function,  
the ratio of intensities
$I(z,\omega)/I^{(0)}(z,\omega)$ can be determined in a way analogous to that in which 
the Purcell factor is determined, as $\kappa_r(\omega)/\kappa_r^{(0)}(\omega)$,
where $\kappa_r(\omega)$ and $\kappa_r^{(0)}(\omega)$ are now expressed through the imaginary
part of the 1D scalar Green’s function. Then, after integrating over the emitting volume
of the structure, as discussed above, one can obtain the ratios
$I(\omega)/I^{(0)}(\omega)$ for a separate $m$-th A layer ($\gamma_m(\omega)$, see Eq. (9)) and
for the whole $N$ layer structure ($\Gamma_N(\omega)$, see Eq. (10)),
which were earlier termed the spontaneous emission intensity enhancement factors.

Here it should be noted that 1D models can satisfactorily describe 
the experimental SE spectra of quasi-one-dimensional structures composed of 
isotropic layers [\onlinecite{Deych,Passias,Levanyuk,Tocci,Dukin,Kuroda,Rybin}].
As an example, there is an approach using the concept of the electromagnetic density
of modes in 1D periodic structures [\onlinecite{Bendickson,Settimi,Fogel,Dowling}], 
that can be exploited in calculations of SE spectra; it is discussed in comparison with the
indirect method from Kirchhoff's law in [\onlinecite{Cornelius}].
Also, for this purpose the scalar formalism based on the Green’s function method 
and the concept of plane scatterers was developed in [\onlinecite{Wubs}] 
and the scattering matrix formalism for the electromagnetic-field S-quantization 
in [\onlinecite{Kaliteevski2015}].
However, only straightforward calculations [\onlinecite{Novotny2,Dowling,Tomas,Dutra,Creatore}] 
taking into account the vectorial nature of the electromagnetic field 
and based on using the dyadic Green's function
can allow a detailed study of the emission from a layered structure, in particular, 
the near-field emission pattern, spectral energy density and LDOS.
Such calculations are based on the so-called direct
method, which is equivalent to the indirect method
[\onlinecite{Wang}] used in the present study. (In this section we focus on a
new aspect of this equivalence related to the replacement of the
dyadic Green's function by the scalar one, which is
possible because \textquotedblleft the LDOS of planar structures is
independent of the polarization states\textquotedblright  [\onlinecite{Wang}].)

Now we will establish a relationship 
between the Kirchhoff’s law approach and Green’s function method. 
For this purpose, we use Eq. (23) given in Appendix, from which one gets
$$
Im{G^{(m)}(z,z)}=\frac{D_0+Re(D_1)\cos2k_az+Im(D_2)\sin2k_az}{2k_a}\:,
$$
$$D_0=\frac{1-|r_Lr_R|^2}{|1-r_Lr_R|^2}\:,\:\:D_1=\frac{r_L+r_R}{1-r_Lr_R}\:,\:\:
D_2=\frac{r_R-r_L}{1-r_Lr_R}\:.
$$
These three functions are related to the 
functions ${\cal F}^{(m)}_{1,2}$, which are proportional to the
field amplitudes outside the photonic crystal,
in the following way:
$$
\frac{2D_0}{1-r^2}=|{\cal F}^{(m)}_1|^2+|{\cal F}^{(m)}_2|^2+|{\cal F}^{(N-m+1)}_1|^2+|{\cal F}^{(N-m+1)}_2|^2\:,
$$
$$Re(D_1)=(1-r^2)Re({\cal F}^{(m)}_1{\cal F}^{(m)*}_2+
{\cal F}^{(N-m+1)}_1{\cal F}^{(N-m+1)*}_2)\:,$$
$$D_2=(1-r^2)({\cal F}^{(m)*}_1{\cal F}^{(m)}_2-
{\cal F}^{(N-m+1)*}_1{\cal F}^{(N-m+1)}_2)\:.$$
Squaring the modulus of the left and right parts of Eq. (13) and
multiplying by the transmission coefficient  $t_{ab}^2$,
one gets the relationship between the total emission intensity and the imaginary part of the Green's
function:
\begin{equation}
n_b(|{\cal{E}}_L^{(m)}(\omega,z)|^2+|{\cal{E}}_R^{(m)}(\omega,z)|^2)=4n_a^2(\omega/c)Im{G^{(m)}(z,z)}\:,
\end{equation}
where $n_b|{\cal{E}}_L^{(m)}(\omega,z)|^2$ and $n_b|{\cal{E}}_R^{(m)}(\omega,z)|^2$ determine
the intensities of the light (emitted at point $z$ of the $m$-th A layer) to the left and
right from the structure.
The last equality expresses the energy conservation law and gives the relation 
between these intensities when using the Green's function method.
Integrating Eq. (20) over the thickness of the $m$-th A layer and
summing over all the layers of the structure and then using Eqs. (2) and (\ref{phi/field}), we get
the following relation
\begin{equation}
\lim_{n''\rightarrow 0}\frac{A_N(\omega)}{n''}= (2\omega/c)^2n_a\sum_{m=1}^N\int_{-a/2}^{a/2}Im{G^{(m)}(z,z,\omega)}dz\:,
\end{equation}
which can be symbolically written as
\begin{equation}
dA(\omega)/dn''\Big|_{n''=0}=2(2\omega/c)^2\int_{(a)} n(z)Im{G(z,z,\omega)}dz\:,
\end{equation}
where the additional factor of $2$ in Eq. (22) in comparison with Eq. (21) is needed because
$A(\omega)$ now is the total absorption coefficient presenting the sum of the absorption coefficients
of light at normal incidence on the left and right sides of the structure,
the subscript $a$ (active) at the integral sign indicates that integration
is over the emitting volume of the structure and the refractive index $n(z)$ is real. 
This consideration, for the sake of shortness and clearness, has been given for a 
periodic structure. The last equation is a generalization of Eq. (21) to an arbitrary case; 
the criteria for the validity of Eq. (22) will be given elsewhere.
(It was verified numerically for various layered structures including non-periodic 
structures with a frequency-dependent refractive index $n(z,\omega)$).
It follows from the equality $\Phi_N(\omega)=\lim_{n''\rightarrow 0}dA_N(\omega)/dn''$ that application of Eq. (22) 
to calculation of the PL spectra (in the case of a spatially homogeneous excitation) is suitable for structures 
with mirror symmetry and for nearly mirror-symmetric ones, e.g., the Fibonacci quasicrystal
(it  becomes symmetric after removal of two outermost layers [\onlinecite{Kaliteevski}],
therefore the many-layer structure can be characterized by a slight violation of 
$\cal{P}$ and $\cal{PT}$-symmetry). 
In particular, using Eqs. (4) and (21) for a single A layer 
(i.e., for symmetric structure) we find that
$$
I_1(\omega) \propto \omega^3
f_T(\omega)\int_{-a/2}^{a/2} Im{G^{(1)}(z,z,\omega)}dz\:,
$$
which is in accordance with Eq. (19), 
as the distribution function $f_T(\omega)$  describes the probability 
of the radiative transitions, which is proportional to $|{\bf d}|^2$
(or, in the quantum approach, to  $\mu^2$).
As was noted above, the function $f_T(\omega)$ is also proportional
to the absorption coefficient, $\alpha(\omega)$, 
hence it follows that there is generally a proportional dependence between 
the absorption coefficient and square of the transition (dipole) moment.
Its determination together with $\alpha(\omega)$ is a 
problem which is to be solved individually for each type of optical transition
associated with a given absorption mechanism.
In practice, the transition moment is a parameter which is
estimated from an experimentally measured spectrum. 
(In the case of forbidden lines the transition moment, for instance,
corresponds to the electric-quadrupole or magnetic-dipole moment operator.)

Thus, the calculation of emission intensity 
based on Eq. (19) and with using the 1D scalar Green's function given by Eq. (23) reduces
to the calculation on the basis of the generalized Kirchhoff's law, see Eq. (4). 
Note that the correspondence between the two approaches 
(for the weak-coupling regime) is related to the use of several analogous
conditions which should be briefly mentioned here:
i) the thermal reservoir or (Markovian) bath and the typical
hierarchy of time-scales 
(in steady-state luminescence experiments),
which is expressed by the cascading inequalities:
\hspace{70 mm}
$\tau_{ph}\ll\omega^{-1}\ll\tau_{th}\ll\tau\ll\tau_{exc}\ll\tau_{res}$,
where $\tau_{ph}\sim l/c$
is the time of light propagation in an atomic system with a dimension $l$
(compare to the value of $\tau_b \sim 10^{-18}$ s in [\onlinecite{Vos}]), 
the vibrational relaxation time $\tau_{th}\sim 10^{-11}-10^{-13}$s,  
the luminescence decay time $\tau \gtrsim 10^{-9}$ s, and 
$\tau_{exc}$ and $\tau_{res}$ are the luminescence excitation time and
the time of change in reservoir temperature, respectively. 
ii) the absence of temporal correlation (the reservoir is memoryless, i.e., 
the coupling of a quantum emitter to the reservoir does not depend on its past,
the memory function is approximated by a delta function)
which is supposed in both the
Langevin approach and Markovian approximation (in the Weisskopf-Wigner theory [\onlinecite{Milonni}]); 
iii) the  small dimensions of the emitting centers compared to the light wavelength, 
that allows one to express the PL intensity in terms of both 
the squared magnitude of the electric field $E(z)$
(in Eq. (12)) and the imaginary part of the Green's tensor (in (Eq. (19), 
where the electric-dipole approximation is exploited); 
iv) a disregard of the stimulated emission
(when deriving Eqs. (3) and (4)) and re-emission 
(which is taken into account by the first-order Maclaurin expansion of
the absorption function $A(\omega)$ in Eq. (2)), and  
correspondingly a disregard of reversible spontaneous emission, 
which is taken into account in the first-order of the perturbation theory by the Fermi’s golden rule
(in the quantum analogue of Eq. (19));
v) the real dielectric function, which allows us to naturally introduce
the PL spectral function  as well as the local density of states.

\newpage

\section{Conclusion}
In conclusion, we have stated the generalized Kirchhoff's law for 
one-dimensional layered structures, which is applicable to the calculation of
the photoluminescence spectra if local thermodynamic equilibrium holds between
the matter and radiation. The overlapping integrals of intensity distributions
of the photonic crystal modes and pump excitation modes are expressed in terms
of the amplitude reflection and transmission coefficients, which is convenient
for numerical calculations and qualitative analysis. The developed approach is
also convenient to obtain the expression for the spontaneous emission intensity 
enhancement factor whose analysis has allowed us to establish some features of 
the modification of the light emission in the case of 1D photonic crystals, in 
particular, the enhancement of emission intensity at the photonic band edges. 
At the same time, the approach using the 1D scalar Green's function
leads to the necessity of calculating the integrals of
the imaginary part of the Green's function and then summing over all emitting layers, 
while the method based on
the generalized Kirchhoff's law allows one to avoid this problem altogether.
The correspondence between these two methods in 
a transparent region was analytically established for a 1D photonic crystal
and confirmed by numerical calculations; in essence,
this approach demonstrates the effective manipulation of 1D Green's function.
Expressions obtained in this paper (Eqs. (12) and (15)) agree with 
the model of independent (incoherent) sources and therefore
can be used for calculating photoluminescence spectra of multi-layered absorbing structures  
while the local thermodynamic condition is met.
As a rule, quasi-one-dimensional structures
allow a much simpler theoretical description of their optical properties 
than two- and three-dimensional systems,
reducing the complexity of the problem and
leading to explicit analytical 
expressions; such is the theory of light emission presented here.
In general, the proposed approach provides the physical basis for luminescence 
engineering of 1D layered structures, in particular, photonic crystals and 
microcavities.
\begin{acknowledgments}
The author greatly thanks E.~L.~Ivchenko and A.~B.~Pevtsov 
for helpful discussions and also M.~Glazov and an anonymous referee for making valuable
comments on the paper.

\end{acknowledgments}

\section*{APPENDIX}

One-dimensional scalar Green's function, $G(z,z',\omega)$, defined for the 
points $z$ and $z'$ of a uniform layer with the thickness $a$ (the both 
coordinates, $z$ and $z'$, are measured from the center of the layer in the same
direction) satisfies the equation [\onlinecite{Ivchenko}] 
$$ (d^2/dz^2 + k_0^2\varepsilon(z))G(z,z',\omega)=-\delta(z,z')\:,$$ 
where the dielectric function $\varepsilon(z)=\varepsilon_a$ for an A layer and 
$\varepsilon(z)=\varepsilon_b$ for a B layer.
For $|z'|\le a/2$,  the Green's function is 
\begin{eqnarray}
&&G(z,z',\omega)=\frac{i}{2k_a}[e^{ik_a|z-z'|}+ \\&& 
\frac{r_Le^{ik_a(z+z')}+r_Re^{-ik_a(z+z')}+2r_Lr_R\cos{k_a(z-z')}}{1-r_Lr_R}]\:,  \nonumber
\end{eqnarray}
where $r_L$ and $r_R$ are the reduced reflection coefficients, on the left and right sides of
the layer, respectively. 
(Equation (23) can be obtained by the standard method for calculating 
the electric field taking into account multiple reflections from the left and right
interfaces and using the expression for the sum of a geometric series.)
In the case of a periodic structure the coefficients $r_L$ and $r_R$ related
to the $m$-th A layer are given by
$$
r_L=e^{i\varphi_a}\frac{r+r_{m-1}e^{i\varphi_b}}{1+rr_{m-1}e^{i\varphi_b}}\:,\:\:
r_R=e^{i\varphi_a}\frac{r+r_{N-m}e^{i\varphi_b}}{1+rr_{N-m}e^{i\varphi_b}}\:.
$$

\end{document}